%% file: ms.tex
\documentclass{article}
\usepackage[utf8]{inputenc}
\usepackage{enumitem}
\usepackage{amsmath}
\usepackage{amsthm}
\usepackage{IEEEtrantools}
\usepackage{tikz}
\usepackage{pgfplots}
\usepackage{caption}
\usepackage{subcaption}
\usepackage{siunitx}
\usepgfplotslibrary{fillbetween}
\usepackage{xspace}
\usepackage{braket}
\usepackage{authblk}
\usepackage[
backend=biber,
style=numeric
]{biblatex}
\renewbibmacro{in:}{}

\pgfplotscreateplotcyclelist{bns}{
    blue,solid\\
    red,densely dashed\\
    brown!60!black,dotted\\
    black,loosely dashed\\
}

\tikzset{
  style0/.style = {violet!70,solid,smooth,semithick},
  style1/.style = {style0,teal,densely dashed},
  style2/.style = {style0,red!90!black,loosely dashed},
  style3/.style = {style0,blue!80,dotted},
  style4/.style = {style0,yellow!70!black,dashdotted},
}

\pgfplotsset{
    compat=newest,
    xmin=0,
    ymin=0,
    width=\linewidth,
    axis y line*=left,
    axis x line*=bottom,
    grid=major,
    grid style={dashed,gray!30,ultra thin},
    xlabel style={font=\footnotesize},
    no markers,
    cycle list name=bns,
    y tick label style={
        /pgf/number format/.cd,
            fixed,
        /tikz/.cd
    },
    legend style={
        cells={anchor=west},
        font=\footnotesize,
        draw=none
    },
    legend columns=3,
    tick label style={font=\scriptsize}
}

\newcommand{\confplot}[6][inference_short]{
    \addplot[style=#6]
    table[x=gibbs_step,y=#2_#3_#4_median,col sep=comma] {data/#1.csv};
    \addlegendentry{#5}
    \addplot+[name path=min,transparent,forget plot]  table[x=gibbs_step,y=#2_#3_#4_min,col sep=comma] {data/#1.csv};
    \addplot+[name path=max,transparent,forget plot] 
    table[x=gibbs_step,y=#2_#3_#4_max,col sep=comma] {data/#1.csv};
    \addplot+[style=#6, very nearly transparent,forget plot] fill between[of=min and max];
}

\newcommand{\pmodel}{p_\text{model}}
\newcommand{\samplefrom}{\sim}
\newcommand{\bas}{BaS\xspace}
\newcommand{\dwave}{\emph{D-Wave Advantage 4.1}\xspace}
\DeclareMathOperator{\given}{\mid}

\theoremstyle{definition}
\newtheorem{definition}{Definition}[section]

\DeclareMathOperator{\precision}{precision}
\DeclareMathOperator{\recall}{recall}
\DeclareMathOperator{\pcdd}{PCDD}
\DeclareMathOperator{\dist}{dist}
\DeclareMathOperator{\med}{MED}

\title{
A hybrid quantum-classical approach for inference on restricted Boltzmann machines
}
\author[1]{M\={a}rti\c{n}\v{s} K\={a}lis\thanks{Corresponding author: martins.kalis.df@lu.lv}}

\author[1]{Andris Loc\={a}ns}

\author[2]{Rolands \v{S}ikovs}

\author[3]{Hassan Naseri\thanks{hassan.naseri@accenture.com}}

\author[1]{Andris Ambainis}

\affil[1]{Center for Quantum Computer Science, Faculty of Computing, University of Latvia}

\affil[2]{Accenture Baltics}

\affil[3]{Accenture United Kingdom}

\date{} 

\begin{document}

\maketitle

\begin{abstract}
Boltzmann machine is a powerful machine learning model with many real-world applications, for example by constructing deep belief networks. %
Statistical inference on a Boltzmann machine can be carried out by sampling from its posterior distribution. %
However, uniform sampling from such a model is not trivial due to an extremely multi-modal distribution. %
Quantum computers have the promise of solving some non-trivial problems in an efficient manner. %
We explored the application of a D-Wave quantum annealer to generate samples from a restricted Boltzmann machine. %
The samples are further improved by Markov chains in a hybrid quantum-classical setup. %
We demonstrated that quantum annealer samples can improve the performance of Gibbs sampling compared to random initialization. %
The hybrid setup is considerably more efficient than a pure classical sampling. %
We also investigated the impact of annealing parameters (temperature) to improve the quality of samples. %
By increasing the amount of classical processing (Gibbs updates) the benefit of quantum annealing vanishes, which may be justified by the limited performance of today's quantum computers compared to classical. %
\end{abstract}

\section{Introduction}

Error-corrected large-scale quantum computers offer asymptotic speed-up for several important problems \cite{grover_fast_1996a, shor_polynomialtime_1997}.
Since these computers are not yet available, there is also a broad interest in the capabilities of noisy intermediate-scale quantum (NISQ) \cite{preskill_quantum_2018} computers.
Although these devices are limited in both scale and fidelity, there is evidence for their advantage over classical devices on some specially designed problems \cite{arute_quantum_2019, mcgeoch_experimental_2013}.
However, a practical advantage of a quantum computer over classical computers is unlikely with current technology.
Therefore, we focus on the strengths of a specific quantum device, the D-Wave quantum annealer, and study a problem that is aligned with those strengths.
This allows us to probe a potential direction where practical benefits of a quantum device might be achieved in the near term.

One machine learning algorithm that maps well to the D-Wave quantum annealer is the restricted Boltzmann machine (RBM).
The performance of the models that include an RBM as a component that is trained using an annealer have been studied in the literature \cite{gircha_training_2021, vinci_path_2020, srivastava_machine_2020, sleeman_hybrid_2020, crawford_reinforcement_2019}.
Gate-based quantum computers have been also used to assist RBM training \cite{zoufal2021variational}.
A more generic model of the quantum RBM includes transverse terms in the energy calculation \cite{adachi_application_2015}.
RBMs have been trained using the D-Wave annealer on the Bars and Stripes (\bas) dataset (images up to $8 \times 8$ pixels) \cite{dixit_training_2021a, li_improved_2020a, benedetti_estimation_2016, xu_adaptive_2021, rocutto_quantum_2021}, down-scaled images or features extracted from MNIST or MNIST-Fashion dataset \cite{sleeman_hybrid_2020, kurowski_applying_2021, li_improved_2020a, adachi_application_2015, koshka_sampling_2020}, and other datasets \cite{dixit_training_2021, srivastava_machine_2020, caldeira_restricted_2020, gardas_quantum_2018}. 
The quality of the models in these papers has been assessed using a mix of measures: their performance on classification problems, image reconstruction, or using some instance-specific figures of merit, such as their likelihood or its approximations.

In this paper, we focus on the classical RBM to highlight the effect of using a quantum annealer.
The RBM is trained classically, then a quantum annealer is used to generate samples from its distribution.
Such sampling has been examined in \cite{koshka_determination_2017, koshka_sampling_2020}.
In particular, we analyze the effect of using samples from a quantum annealer to initialize Markov chains used to generate new samples from the RBM distribution.
\cite{mazzanti_efficient_2020} consider several classical initialization techniques for Markov chains to improve sampling from the model probability distribution of an RBM.
We build on these papers by considering substantially larger models and a larger dataset (\bas with $12 \times 12$ images) that is  more representative of practical problems.
To validate the findings, the experiments are also run on a smaller Labeled Shifter Ensemble dataset \cite{schulz_investigating_2010} .
We use a fraction of the positive images (with either only bars or only stripes) to train RBMs, which allows us to analyze how introducing quantum annealing changes the generalization performance of the trained models, i.e., how well the models generate positive images that are not present in the training set.
We used highly informative figures of merit to evaluate and report the findings.

Compared to supervised classification problems with reliable ground truth, evaluating models on generative problems can be difficult.
Although all evaluation metrics have their pros and cons \cite{borji_pros_2022}, a combination of precision and recall offers a highly interpretable overview of model performance and differentiates between  failure cases.
For many practical generative problems defining and calculating precision and recall can be difficult, and, instead of two separate scores, the trade-off between the two is used \cite{sajjadi2018assessing, kynkaanniemi2019improved}.
However, most machine learning problems that can be approached by current quantum hardware are still of limited size and with more structure than most problems tested in classical machine learning.
These two aspects allow us to use a simpler and more interpretable approach to precision and recall to study the performance of quantum devices on machine learning problems.
Both of these properties afford the use of meaningful performance measures of generative models to assess whether they merely compress and memorize the training dataset, or do they represent some underlying structure that leads to correct generated samples outside the training dataset.

The main contributions of this paper can be summarized as follows:
\begin{enumerate}
    \item We push the limits of quantum machine learning by implementing an RBM model larger than any other in the literature for the datasets considered. 
    These novel results shed light on the scalability of quantum annealing for machine learning models as the problem size increases.
    \item We propose a hybrid approach of using a quantum annealer to initialize classical Markov chains that produce samples from a distribution. 
    This hybrid approach outperforms pure quantum or classical compute methods.
    \item We use more informative figures of merit compared to prior studies to assess the effects of applying quantum annealing to sample generation for a machine learning model.
\end{enumerate}

The rest of the paper is organized as follows.
Section~\ref{sec:rbm} defines the RBM model used in experiments.
Section~\ref{sec:qa} introduces quantum annealing concepts relevant to the paper.
Section~\ref{sec:setup} describes the setup of the main experiments and introduces the figures of merit used to analyze the results.
Section~\ref{sec:results} presents the results and analysis of the experiments.
Section~\ref{sec:conclusion} discusses the larger implications of the results, as well as outlines possible future research directions.

%%%%%%%%%%%%%%%%%%%%%%%%%%%%%%%%%%%%%%%%%%%%%%%%%%%%%%%%%%%%%
\section{Restricted Boltzmann Machine}
\label{sec:rbm}
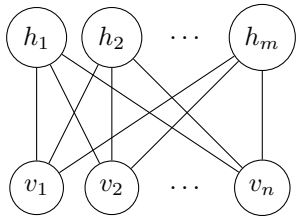
\begin{figure}
    \centering
    \begin{tikzpicture}
    
    % Visible layer
    \node[circle,draw] (h1) at (0,2) {$h_1$};
    \node[circle,draw] (h2) at (1,2) {$h_2$};
    \node[] at (2,2) {$\dots$};
    \node[circle,draw] (hm) at (3,2) {$h_m$};
    
    % Hidden layer
    \node[circle,draw] (v1) at (0,0) {$v_1$};
    \node[circle,draw] (v2) at (1,0) {$v_2$};
    \node[] at (2,0) {$\dots$};
    \node[circle,draw] (vn) at (3,0) {$v_n$};
    
    % Connections
    \draw (v1) -- (h1);
    \draw (v1) -- (h2);
    \draw (v1) -- (hm);
    \draw (v2) -- (h1);
    \draw (v2) -- (h2);
    \draw (v2) -- (hm);
    \draw (vn) -- (h1);
    \draw (vn) -- (h2);
    \draw (vn) -- (hm);
    \end{tikzpicture}
    \caption{Diagram of an RBM with $n$ visible units and $m$ hidden units.}
    \label{fig:rbm}
\end{figure}

RBM (Figure~\ref{fig:rbm}) is a machine learning model that aims to learn a probability distribution that represents the input data \cite{hinton2012practical}. 
It has a structure of a visible and a hidden layer consisting of binary-valued nodes ${v = (v_1, v_2, \ldots, v_n)}$ and ${h = (h_1, h_2, \ldots, h_m)}$ respectively.
All the visible nodes are connected with all hidden nodes.
There are no connections within the layers.

The probability distribution is learned using the energy concept:
\begin{align}
    E(v,h) = -\sum_{i=1}^n{a_i v_i} - \sum_{j=1}^m{b_j h_j} -\sum_{i=1}^n{\sum_{j=1}^m{w_{ij} v_i h_j}},
\end{align}
where $a = (a_1, a_2, ..., a_n)$ are biases for the visible nodes, $b = (b_1, b_2, ..., b_m)$ are biases for the hidden nodes and $w = (w_{ij})$ is the upper-triangular weight matrix.
The probability $\pmodel(v, h)$ of finding a configuration $(v,h)$ at temperature $T$ is equal to $\pmodel(v,h) = \frac{e^{-\beta E(v,h)}}{Z}$, where the partition function $Z = \sum_{v,h}{e^{-\beta E(v,h)}}$ is the sum over all the possible configurations and $\beta = 1/T$.
$T$ is typically set to be equal to $1$ for classical training, but can be used to adjust the model to the effective temperature of the quantum annealer.
The goal of training an RBM is to maximize the likelihood of the model parameters given data, i.e., maximize $L(a, b, w \given S) = \prod_{v \in S} \pmodel(v \mid a, b, w)$ over the choice of $a, b, w$, where $S$ is the set of training samples.

Likelihood maximization is done using stochastic gradient descent.
A batch of samples is used to approximate the derivative of the negative log-likelihood function with respect to each weight $w_{ij}$, visible bias $a_i$, and hidden bias $b_j$.
Weights and biases are then updated using the derivative scaled by the learning rate, mostly following the training scheme described in \cite{hinton2012practical}.
When updating the weights of an RBM, one can distinguish between the positive and the negative phases given by:
\begin{align}
    \Delta w_{ij} \propto \overbrace{\langle v_ih_j \rangle_{\text{fixed data}}}^{\text{positive phase}} - \overbrace{\langle v_i h_j \rangle_{\text{all configurations}}}^{\text{negative phase}},
\end{align}
where the positive phase term is averaged over configurations with the visible unit values fixed to the training data, and the negative phase term is averaged over all possible configurations.
The positive phase term helps the model lower the energy of the observed samples from the training set and the negative phase helps raise the energy of the states that have no connection to the data but might have low energy otherwise.

Sampling from the model distribution, necessary for the negative phase of RBM training, can be time and resource consuming, because of the large number of possible configurations in the partition function calculation.
The classical heuristics for approximate sampling from the RBM model distribution are typically based on Gibbs sampling.
$k$-th full Gibbs update is the sampling of the hidden layer values given fixed visible layer values (${h^{(k)} \samplefrom \pmodel(h \given v^{(k)})}$) and back again (${v^{(k+1)} \samplefrom \pmodel(v \given h^{(k)})}$) for $k>0$.

If the visible layer values are fixed, the probabilities from which hidden layer values are sampled are:
\begin{align}
	\pmodel(h^{(k)}_j = 1 \given v^{(k)}) = \frac{1}{1 + \exp{(-\sum_i{w_{ij} v^{(k)}_i}-b_j})},\\
	\pmodel(v^{(k+1)}_i = 1 \given h^{(k)}) = \frac{1}{1 + \exp{(-\sum_j{w_{ij} h^{(k)}_j}-a_i})}.
\end{align}

The main difference for the samplers used in this paper is the initial values for the visible units ($v^{(0)}$).

\begin{description}
    \item[Naive Gibbs sampling]
		In the ``naive'' approach, we start with random binary vectors for the visible layer.
		A large number of Gibbs updates may be necessary to ensure that the resulting samples represent the model distribution well.
    \item[Contrastive Divergence (CD)]
		The initial values for the visible units in CD training are the training samples.
        In practice, a low number of Gibbs updates is used for training.
        Often just one full Gibbs update is used.
        This is referred to as CD-1 training.
		This means that the samples are likely to be more concentrated around the training data points, if the number of Gibbs updates is not large.
		It is possible that the model distribution still has high probabilities for unrepresentative points far from the training data points.
		CD often works well in practice.
\end{description}

%%%%%%%%%%%%%%%%%%%%%%%%%%%%%%%%%%%%%%%%%%%%%%%%%%%%%%%%%%%%%
\section{Quantum annealing}
\label{sec:qa}

Idealized quantum annealing can be described in terms of a time-dependent Hamiltonian $H(s)$:

\begin{align}
	H(s) &= -A(s)\sum_i{\sigma_i^x} + B(s)H_\text{target},\\
	H_\text{target} &= \sum_i{h_i\sigma_i^z} + \sum_{i<j}{J_{ij}\sigma_i^z\sigma_j^z},
\end{align}
where $H_\text{target}$ is the target Hamiltonian encoding the problem of interest, $s=t/t_a$; $t$ is time, $t_a$ is the annealing time, $h_i$ and $J_{ij}$ are tunable dimensionless parameters, and $A(s)$ and $B(s)$ are monotonic functions such that $A(0) \gg B(0) \approx 0$ and ${B(1) \gg A(1) \approx 0}$, and $\sigma_i^x$, $\sigma_i^z$ are Pauli $X$ and $Z$ operators acting on qubit $i$ with the transformation $\begin{bmatrix}0&1\\1&0\end{bmatrix}$ and $\begin{bmatrix}1&0\\0&-1\end{bmatrix}$ respectively \cite{amin_searching_2015}.

Quantum annealing is closely related to adiabatic computing.
In adiabatic computing, if (a) the quantum state $|\psi(0)\rangle$ is initialized to an easily prepared ground state\footnote{Eigenstate of the Hamiltonian with the lowest eigenvalue.}, (b) $H(s)$ has a unique ground state for all $s \in [0,1]$, and (c) the evolution from $H(0)$ to $H(1)$ is sufficiently slow, the final state of the qubits $|\psi(1)\rangle$ will correspond to the ground state of $H(1)$.\cite{farhi_quantum_2000}
This can be useful for optimization problems, where the aim is to find a state that minimizes the energy of the system.
In this case, only the state with the lowest energy returned by the annealer might be of interest, and other returned states can be discarded.

It is also possible to use such an annealer for sampling applications.
% Not sure if I need to go into more detail here.
Samples from an ideal quantum annealer would follow the Boltzmann distribution at a tunable temperature.
In practice, due to various physical limitations and noise, samples from the annealer do not truly correspond to any $H(s), s \in [0,1]$.
However, they may come from a distribution that is close enough to the target distribution to be useful.

There are several restrictions that one needs to overcome to use a quantum annealer for either of these two applications.
First, the problem needs to be embedded on a chip with a limited number of qubits and connections between them.
While the RBM structure maps well to the architecture of the D-Wave annealers, and \dwave has more than \num{5000} qubits, the actual number of RBM units that can be embedded on the annealer is far lower\footnote{The highest largest fully connected RBM we have embedded on \dwave has \num{168} visible units and \num{160} hidden units.}.
The connection strengths between qubits also has a specific range, adding further limits on RBMs that can be embedded on the device.
See \cite{pochart_challenges_2022} for a more detailed overview of the challenges of using D-Wave annealers to sample from a Boltzmann distribution.

%%%%%%%%%%%%%%%%%%%%%%%%%%%%%%%%%%%%%%%%%%%%%%%%%%%%%%%%%%%%%
\section{Experimental setup}
\label{sec:setup}

To compare the properties of classical and quantum annealing-based initialization for sampling from a distribution of images using Markov chain Monte Carlo, we use the \bas dataset.
We perform additional tests using the Labeled Shifter Ensemble \cite{schulz_investigating_2010}, described in more detail in Appendix~\ref{sec:shifter}.
\bas consists of all $n \times n$ binary pixel images, where either only entire stripes or only entire bars are colored (i.e., have value~\num{1}).
\bas is a commonly used dataset in quantum machine learning research (e.g. \cite{dixit_training_2021a, zeng2019learning}) due to its structure and adjustable size.
Like real-world images, positive \bas images (those with either only bars or only stripes colored) are embedded in a substantially larger space of all $n \times n$ binary images.
For $n \times n$ images there are $2^{n^2}$ total images and $2 \cdot 2 ^ n - 2$ positive images.
Unlike real-world images, the \bas dataset allows us to calculate some properties of the images that are useful in result analysis, such as the exact probability of each image in the true data distribution, as well as the distance of a negative image to its closest positive image. 

The maximum size of an RBM with $n^2$ visible and $n^2$ hidden units (architecture that maps particularly well to the annealer) that we could embed on the \dwave was for $n=12$, so we used images of size $12 \times 12$ for experiments in this paper.
This more than doubles the number of pixels compared to the previous largest \bas experiments on D-Wave devices in the literature that we know of.
This size still allows us not to use feature reduction to represent the images with RBMs, ensuring that features and their relations typical to images remain in the training dataset.
The number of all positive images for this dataset is \num{8190}.

Two types of RBMs were trained classically using custom-built RBM classes.
One used Naive Gibbs sampling with \num{50} full Gibbs updates for estimating the negative phase; the other used contrastive divergence with one full Gibbs update (CD-1).
The former was expected to perform better on the tests with classical random initialization since it matches the training procedure.
We tested the CD-1 approach due to its widespread use in RBM training.
For each type of RBM, five identical models were initialized with random weights and biases.
Five training datasets of \num{512} positive \bas images (approximately \SI{6.25}{\percent} of all positive images) were i.i.d. sampled from all positive $12 \times 12$ \bas images.
Standard rather than stochastic gradient descent was used for training, i.e., the batch size was set to \num{512}~--- the full training dataset.
This was done to emulate parameters that we would use when training RBMs using a quantum annealer for initialization in negative phase estimation since a large batch size minimizes the required number of experiments on the quantum annealer.
One Naive and one CD-1 RBM was trained on each of the datasets.

For each model and initialization configuration, \num{40000} Markov chains with \num{144} binary random variables were created.
The chains were initialized using either \emph{classical} or \emph{D-Wave} strategy.
\emph{Classical} initialization used pseudo-random binary strings.
\emph{D-Wave} initialization used samples from the D-Wave annealer simulating the RBM at temperatures $T \in \set{8, 16, 32, 64}$.
To improve the quality of the D-Wave generated samples, two parameters were changed from their default values.
Auto-scaling of weights and biases to use the whole range of acceptable device values was turned off to see the effects of choosing different temperatures.
The sampler was set to use ten spin-reversal transforms\footnote{In a problem formulation with RBM unit values $\{-1, +1\}$ this involves flipping the sign of a subset of units and their associated weights and biases. The results returned are already translated to correspond to the original problem formulation.} to account for some of the systematic errors of the annealer.
A separate set of Markov chains was used for each of the temperatures $T$.

For each of the Markov chains, \num{1000} full Gibbs updates were run.
Figure~\ref{fig:initializations} shows some examples of the initial and final states of the Markov chains using different initialization strategies.
For a subset of the models, \num{10000} full Gibbs updates were run to test the properties of longer runs.
For the first \num{100} updates the figures of merit (defined below) were calculated after every update.
For the rest of the updates they were calculated after every \num{10} updates.

\input{figures/patterns.tex}

Results for various combinations of the type of RBM (Naive, CD-1), initialization (classical, D-Wave) and temperature $T$ ($T=1$ for classical initialization, $T \in \set{8, 16, 32, 64})$ for D-Wave initialization) are graphed in Section~\ref{sec:results}.
For each combination graphed, the line corresponds to the median value of the figure of merit at each of the Gibbs update number.
The area between the minimum and the maximum of the five values is shaded in the same color.
For the long-run (\num{100000} Gibbs updates) experiments, only one model is tested for each combination, and only one line is plotted.

The two main figures of merit used to assess the properties of classical and D-Wave initializations are $\precision$ (Definition~\ref{def:precision}) and $\recall$ (Definition~\ref{def:recall}.
\begin{definition}[Precision]\label{def:precision}
Given a multiset $G$ of generated samples and a set $X$ of positive samples, 
$\precision(G, X)$ is $$\precision(G, X) = \frac{|\set{x \in G \given x \in X}|}{|G|}.$$
\end{definition}

\begin{definition}[Recall]\label{def:recall}
Given a multiset $G$ of generated samples and a set $X$ of positive samples,
$\recall(G, X)$ is $$\recall(G, X) = \frac{|\set{x \in X \given x \in G]}|}{|X|}.$$
\end{definition}

Note that recall as defined here is dependent on the number of generated samples.
A larger set of generated samples is expected to contain more of the low-probability positive images.
In all tests we generate \num{40000} samples (with \num{8190} positive images), thus all the recall figures in this paper are comparable.

Two further figures of merit provide additional insights: the positive-case distribution distance ($\pcdd$, Definition~\ref{def:pcdd}) and negative-case mean edit distance ($\med$, Definition~\ref{def:med}).
For many problems, we might be able to recognize positive examples and discard the negative examples.
$\pcdd$ shows how well the remaining positive examples represent the true distribution.
$\med$ offers insight into the properties of the negative examples.
For the \bas dataset, $\med$ can show whether the negative examples are just stochastic deviations by a small number of pixels from the positive examples that might be corrected in another step (low $\med$), or if examples are typically stuck in some local energy minima that are further away from any positive examples (high $\med$).
Note that calculating $\med$ might be computationally expensive for most datasets.

\begin{definition}[Positive-case distribution distance]\label{def:pcdd}
	Given a multiset $G$ of generated images and a set $X$ of positive images,
	$\pcdd$ is $$\pcdd(G, X) = \frac{\sum_{x \in X}{|\set{g \in G \given g=x}|}}{|X|}.$$
\end{definition}

\begin{definition}[Edit distance]
	Given an image $g$ and a set $X$ of positive images,
	The edit distance $\dist(g, X)$ is the minimum number of pixels that need to be changed in image $g$ to produce a positive image $x \in X$.
\end{definition}

\begin{definition}[Negative-case mean edit distance ($\med$)]\label{def:med}
	Given a multiset $G$ of generated images and a set $X$ of positive images,
	the negative-case mean edit distance $\med$ is $$\med(G, X) = \frac{\sum_{g \in G^-}{\dist(g, X)}}{|G|},$$
	where $G^- = \set{g \in G \given g \notin X}$.
\end{definition}

%%%%%%%%%%%%%%%%%%%%%%%%%%%%%%%%%%%%%%%%%%%%%%%%%%%%%%%%%%%%%
\section{Results}
\label{sec:results}

\input{figures/naivevcd1.tex}

\input{figures/cd1.tex}

Although models trained with the Naive approach perform better than CD-1 trained models on the \bas dataset in our experiments (Figure~\ref{fig:naive_v_cd1}), the benefits of D-Wave initialization are bigger for the CD-1 trained RBMs (Figure~\ref{fig:cd1}).
This is likely due to spurious local energy minima far away from any data points that the CD-based RBM training is not designed to learn to correct.
A randomly initialized Markov chain is likely to encounter some of these spurious minima.
While their energy might be higher than that of positive samples, they do slow down the mixing of the chains.
Initializing the chains closer to the lower-energy minima using samples from D-Wave can help avoid some of the spurious minima.
The additional experiments described in Appendix~\ref{sec:shifter} show that the benefits of D-Wave initialization for the CD-1 trained RBMs are less pronounced in the smaller \num{19} bit Shifter problem, where the CD-1 based training is able to explore more of the space. 
Figure~\ref{fig:naive_v_cd1} shows that D-Wave initialization outperforms classical initialization on both precision and recall for up to \num{1000} Gibbs updates, and figure \ref{fig:long} suggests the effect lasts for up to \num{100000} Gibbs updates, and possibly more.

\input{figures/classicvdw.tex}

\input{figures/longexperiments.tex}

While current quantum annealers cannot generate a very accurate sample from an arbitrary distribution, they perform better than sampling from the uniform distribution in most cases.
Consequently, D-Wave generated initial images are closer to some positive image than uniformly pseudo-randomly initiated images, and fewer Gibbs updates are necessary for them to be transformed into a positive image.
This is illustrated in Figure~\ref{fig:classic_v_dw_precision}.

For some use-cases, starting too close to a small number of positive datapoints might be disadvantageous.
However, adjusting the temperature of the embedding can help in these cases.
Figure~\ref{fig:classic_v_dw_recall} suggests that at $T=8$ more of the samples generated are already close to some local minima in the energy landscape, and Markov chains initialized using these samples remain close to them, and are unable to explore the full state space.
This is supported by Figure~\ref{fig:top10}, which shows a substantially higher concentration of top-10 most produced positive samples for initializations at lower temperatures.
Classical initialization produces the lowest concentration around top-10 positive samples.

\input{figures/top10.tex}

As noted in \cite{koshka_comparison_2020}, quantum annealing and classical methods can produce samples of similar quality, yet from different distributions.
This allows one to combine the two methods to improve results.
In our experiments, initial states sampled uniformly from the union of classical random initial states and D-Wave samples at $T=8$, outperformed classical initialization at recall and precision at a small number of Gibbs updates (Figure~\ref{fig:classic_dw_hybrid}). 
A combination of classical initialization and D-Wave samples at $T=32$ offers a slight improvement over pure D-Wave samples at $T=32$.
However, the benefit was less pronounced, since sampling at higher temperatures already ensures that samples come from a distribution that is closer to the uniform, resembling the hybrid approach of combining classical random samples and D-Wave samples at lower temperatures.
For example, D-Wave initialization at $T=64$ delivers very similar results to hybrid initialization at $T=32$.

\input{figures/classic_dw_hybrid.tex}

For large sample counts, the per-sample processing time accounts for the majority of the total QPU access time.
For the D-Wave annealer this consists of a user-specified anneal and delay (pause between samples) time, and hardware-dependent readout times.
The samples used in this experiment were produced using the default linear anneal over \SI{20}{\us}, \SI{20}{\us} delay, and \SI{214}{\us} for readout.
Processing \num{10000} samples thus took \SI{2.54}{\second}.
For reference, one full Gibbs update for \num{10000} samples implemented in Python using the NumPy library takes approximately \SI{110}{\ms} on a {\SI{1.2}{\GHz} Dual-Core Intel Core m5} processor.
Thus \num{23} Gibbs updates could be performed during the time it takes the D-Wave annealer to produce \num{10000} samples for the models used in the experiment.
Note that this comparison is hardware-specific for both technologies.

\section{Conclusions}
\label{sec:conclusion}

Boltzmann machine is a powerful machine learning model.
We demonstrated that quantum annealer samples can improve sample generation from a restricted Boltzmann machine compared to random initialization.
The hybrid approach outperforms pure quantum or classical compute methods.
We used the highly structured \bas dataset of a size not yet tested on a quantum annealer.
These novel results shed light on the scalability of quantum annealing for machine learning models as the problem size increases.

For this particular problem, a pure quantum annealing approach at the moment does not offer a sampling speed-up advantage.
However, sampling at sufficiently high temperatures or combining the classical and quantum annealing approaches to initialization offers a slight advantage in the quality of the samples produced at some number of Gibbs updates.
For example, if we are interested in producing samples with the highest level of recall, a hybrid initialization produces a higher peak recall, as well as higher precision at the same number of steps, compared to the classical random initialization.

The results and methods presented in the paper can help guide further use of quantum annealing in training RBMs and other generative models that rely on MCMC training.
While the current guidance is to perform some Gibbs updates after initializing Markov chains with samples from a quantum annealer to get the sample distribution closer to the Boltzmann distribution, precision and recall (or another sample diversity measure when calculating recall is unfeasible) are concrete tools that can be used to determine the number of updates necessary.
With the substantial performance differences between Naive and CD-1 trained RBMs, it would also be interesting to train RBMs with quantum annealing samples for MCMC initialization in the negative phase of the training.

\section*{Acknowledgments}

The study has been supported by Accenture Baltics and conducted in collaboration with a dispersed Accenture team under an ``Agreement on quantum computing use case research'' (March, 2021). We thank Carl M. Dukatz (Accenture USA), Shreyas Ramesh (Accenture USA), Tekin E. Ozmermer (Accenture Baltics), Zane Kalnina (Accenture Baltics) and Karlis Freivalds (University of Latvia) for their contributions throughout the study.

\printbibliography

\clearpage
\appendix

\section{The Labeled Shifter Ensamble}
\label{sec:shifter}

\subsection{The Definition of the Ensamble}
\begin{figure}[h]
\centering
\includegraphics[scale=0.2]{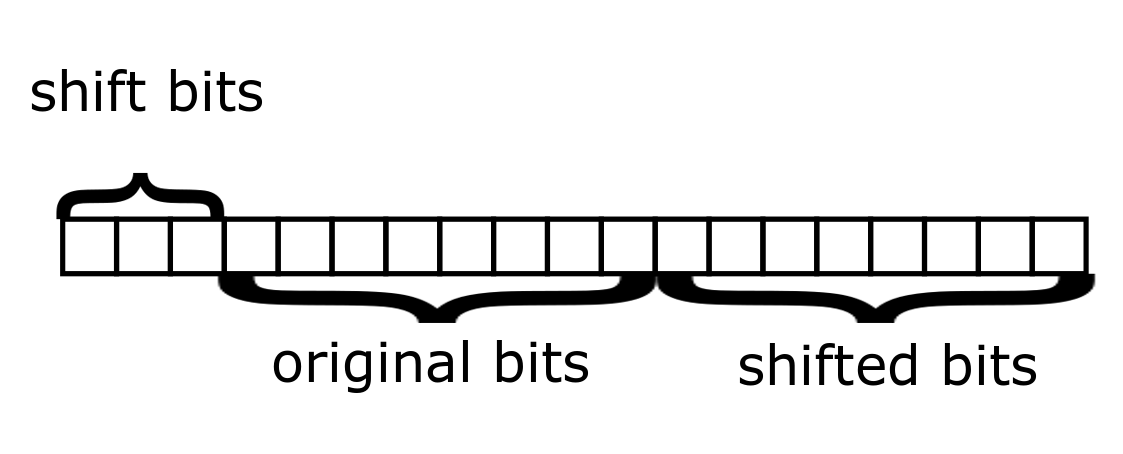}
\caption{The structure of the shifter sample}
\label{fig:shifter}
\end{figure}

A positive sample of the Shifter ensemble (or, equivalently, the Shifter problem) is defined as follows. It has $2n+3$ bits where  $n$ is the number of the original bits as seen in Figure~\ref{fig:shifter}. 

The first three bits determine whether those original bits should be shifted one position to the left, not shifted at all, or shifted one position to the right. Then the original $n$ bits follow and the last $n$ bits are the original bits shifted accordingly to the first three bits.

\subsection{Experimental Setup}

In the experiment, we used a Shifter ensamble with $n=8$ and $3 \cdot 2^8 = 768$ positive examples. Five different RBM models were trained using the contrastive divergence sampler and five models using the naive Gibbs sampler (see Section~\ref{sec:rbm}). Note that any bitstring of length \num{19} can be turned into a positive example of the Shifter problem with no more than eight bit flips, while a $12 \times 12$ randomly initialized binary image typically requires more than \num{50} changes (Figure~\ref{fig:classic_v_dw_mad}) to reach a positive Bars and Stripes (\bas) example. This makes the contrastive divergence sampler more effective at exploring the whole \num{19} bit string space, compared to $12 \times 12$ binary image space.

For each model, a different set of \num{256} randomly selected positive Shifter samples were used. Each model was trained with the stochastic gradient descent method taking \num{256} as the batch size which made it equivalent to simple gradient descent. And each model was trained for \num{4000} epochs.

Using the trained models \num{4000} samples of size \num{19} were generated using classical random initialization or D-Wave samplers with temperatures $T \in \{2,4,8,64\}$ for the CD-trained RBMs or $T \in \{2,3,4,5,6,7,8,64\}$ for the RBMs trained with the naive sampler.

To evaluate the generated sample sets each sample was updated with Gibbs updates \num{1000} times and after every update, two figures of merit were calculated~--- precision (Definition~\ref{def:precision}) and recall (Definition~\ref{def:recall}).

\subsection{Results}

Similarly to \bas the quantum-assisted sample sets had a higher precision when trained with the naive model (Figure~\ref{fig:shifter_naive}). The highest precision was reached with D-Wave temperature $T = 4$. 

\input{figures/shifter_naive.tex}

At low temperatures, the RBM weights that needed to be embedded in the D-Wave annealer started competing with the coupler strengths used to ensure the consistency of the physical qubit values within the qubit chains forming the logical qubits. For example, at $T=4$, only about \SI{0.5}{\percent} chains had inconsistent qubit values; at $T=2$ approximately \SI{5}{\percent} chains had inconsistent qubit values. We hypothesize that this is the reason why precision started falling for RBM temperatures below $T=4$ (see Figure~\ref{fig:shifter_naive_small_t_precision}).

\input{figures/shifter_naive_small_T.tex}

All the sample sets from the CD-1 trained models had a similar precision after just a few Gibbs updates (Figure~\ref{fig:shifter_cd1}). Initializing the samples close to high probability states substantially reduces the maximum recall achieved.

\input{figures/shifter_cd1.tex}

\end{document}

%% file: figures/patterns.tex
\pgfplotstableread{data/patterns.txt}{\patterns}

\newcommand{\pattern}[1]{
  \begin{tikzpicture}[scale=0.0875]
      \foreach \imnum in {0, ..., 9} {
        \foreach \x in {0, ..., 11} {
          \foreach \y in {0, ..., 11} {
            \pgfmathsetmacro{\row}{1440 * #1 + 144*\imnum + 12*\x+\y}
            \begin{scope}[shift={(14*\imnum,0)}]
              \pgfplotstablegetelem{\row}{fill}\of\patterns
              \ifthenelse{1 = \pgfplotsretval}{
                  \filldraw [fill=gray, draw=lightgray, line width=0.01mm] (\x,\y) rectangle (\x+1,\y+1);
              }{
                  \draw [color=lightgray, line width=0.01mm] (\x,\y) rectangle (\x+1,\y+1);
              }
            \end{scope}
          }
        }
      }
  \end{tikzpicture}
}

\begin{figure}
  \begin{center}
    \begin{subfigure}{\textwidth}
      \pattern{0}\\[-3ex]
      \caption{Classical initialization.\bigskip}\label{fig:initializations_classical}
    \end{subfigure}
    \bigskip
    \begin{subfigure}{\textwidth}
      \pattern{1}\\[-3ex]
      \caption{Classical initialization after \num{1000} Gibbs updates with naively-trained RBM.}\label{fig:initializations_classical_result}
    \end{subfigure}
    \bigskip
    \begin{subfigure}{\textwidth}
      \pattern{2}\\[-3ex]
      \caption{D-Wave initialization ($T=8$) for a naively-trained RBM.}\label{fig:initializations_dwave}
    \end{subfigure}
    \bigskip
    \begin{subfigure}{\textwidth}
      \pattern{3}\\[-3ex]
      \caption{D-Wave initialization ($T=8$) after \num{1000} Gibbs updates with naively-trained RBM.}\label{fig:initializations_dwave_results}
    \end{subfigure}
    \bigskip
    \begin{subfigure}{\textwidth}
      \pattern{4}\\[-3ex]
      \caption{D-Wave initialization ($T=8$) for a CD-1-trained RBM.}\label{fig:initializations_cd}
    \end{subfigure}
    \bigskip
    \begin{subfigure}{\textwidth}
      \pattern{5}\\[-3ex]
      \caption{D-Wave initialization ($T=8$) after \num{1000} Gibbs updates with CD-1-trained RBM.}\label{fig:initializations_cd_results}
    \end{subfigure}
    
    \caption{Types of initialization and results produced after \num{1000} Gibbs updates. Each image has $12 \times 12$ binary pixels, corresponding to the \num{144} visible units of the RBMs. Shaded pixels correspond to the value \num{1}, white pixels correspond to value \num{0}. The resulting images in (b), (d), (f) correspond to the initializations shown in (a), (c), (e).}
    \label{fig:initializations}
  \end{center}
\end{figure}

%% file: figures/naivevcd1.tex
\begin{figure}
  \begin{center}
  \ref*{n_v_cd1_legend}
  \begin{subfigure}[b]{0.49\textwidth}
    \begin{tikzpicture}
      \begin{semilogxaxis}[
            xlabel=Gibbs update,
            xmax=1000,
            xmin=1,
            legend pos=south east,
            legend to name=n_v_cd1_legend
        ]
        \confplot{Naive}{classic}{precision}{Naive, classical}{style0}
        \confplot{Naive}{dwave_64}{precision}{Naive, D-Wave, $T=64$}{style1}
        \confplot{CD1}{classic}{precision}{CD-1, classical}{style2}
        \confplot{CD1}{dwave_8}{precision}{CD-1, D-Wave, $T=8$}{style3}
        
      \end{semilogxaxis}
    \end{tikzpicture}
    \caption{Precision}\label{fig:naive_v_cd1_precision}
  \end{subfigure}
  \begin{subfigure}[b]{0.49\textwidth}
    \begin{tikzpicture}
      \begin{semilogxaxis}[
            xlabel=Gibbs update,
            xmax=1000,
            xmin=1,
        ]
        \confplot{Naive}{classic}{recall}{Naive, classical}{style0}
        \confplot{Naive}{dwave_64}{recall}{Naive, D-Wave, $T=64$}{style1}
        \confplot{CD1}{classic}{recall}{CD-1, classical}{style2}
        \confplot{CD1}{dwave_8}{recall}{CD-1, D-Wave, $T=8$}{style3}
        \legend{}
      \end{semilogxaxis}
    \end{tikzpicture}
    \caption{Recall}\label{fig:naive_v_cd1_recall}
  \end{subfigure}
  \caption{RBMs trained with the naive approach outperform CD-1-trained RBMs on both precision and recall. For both models the best-performing temperature was selected for D-Wave initialization.}
  \label{fig:naive_v_cd1}
  \end{center}
\end{figure}

%% file: figures/cd1.tex
\begin{figure}
  \begin{center}
  \ref*{cd1_legend}
  \begin{subfigure}[b]{0.49\textwidth}
    \begin{tikzpicture}
      \begin{semilogxaxis}[
            xlabel=Gibbs update,
            xmax=1000,
            xmin=1,
            legend pos=south east,
            legend to name=cd1_legend
        ]
        \confplot{CD1}{classic}{precision}{CD-1, classical}{style0}
        \confplot{CD1}{dwave_8}{precision}{CD-1, D-Wave, $T=8$}{style1}
        \confplot{CD1}{dwave_16}{precision}{CD-1, D-Wave, $T=16$}{style2}
        \confplot{CD1}{dwave_32}{precision}{CD-1, D-Wave, $T=32$}{style3}
        \confplot{CD1}{dwave_64}{precision}{CD-1, D-Wave, $T=64$}{style4}
        
      \end{semilogxaxis}
    \end{tikzpicture}
    \caption{Precision}\label{fig:cd1_precision}
  \end{subfigure}
  \begin{subfigure}[b]{0.49\textwidth}
    \begin{tikzpicture}
      \begin{semilogxaxis}[
            xlabel=Gibbs update,
            xmax=1000,
            xmin=1,
        ]
        \confplot{CD1}{classic}{recall}{CD-1, classical}{style0}
        \confplot{CD1}{dwave_8}{recall}{CD-1, D-Wave, $T=8$}{style1}
        \confplot{CD1}{dwave_16}{recall}{CD-1, D-Wave, $T=16$}{style2}
        \confplot{CD1}{dwave_32}{recall}{CD-1, D-Wave, $T=32$}{style3}
        \confplot{CD1}{dwave_64}{recall}{CD-1, D-Wave, $T=64$}{style4}
        \legend{}
      \end{semilogxaxis}
    \end{tikzpicture}
    \caption{Recall}\label{fig:cd1_recall}
  \end{subfigure}
  \caption{CD-1 inference with various initializations. CD-1 trained models benefit from initialization closer to positive images. D-Wave initializations outperform classical random initialization, and lower temperatures outperform higher ones (for the temperatures tested).}
  \label{fig:cd1}
  \end{center}
\end{figure}

%% file: figures/classicvdw.tex
\begin{figure}
  \begin{center}
  \ref*{init_legend}
  \begin{subfigure}[b]{0.49\textwidth}
    \begin{tikzpicture}
      \begin{semilogxaxis}[
            xlabel=Gibbs update,
            xmax=1000,
            xmin=1,
            legend pos=south east,
            legend to name=init_legend
        ]
        \confplot{Naive}{classic}{precision}{classical}{style0}
        \confplot{Naive}{dwave_8}{precision}{D-Wave, $T=8$}{style1}
        \confplot{Naive}{dwave_16}{precision}{D-Wave, $T=16$}{style2}
        \confplot{Naive}{dwave_32}{precision}{D-Wave, $T=32$}{style3}
        \confplot{Naive}{dwave_64}{precision}{D-Wave, $T=64$}{style4}
      \end{semilogxaxis}
    \end{tikzpicture}
    \caption{Precision}\label{fig:classic_v_dw_precision}
  \end{subfigure}
  \begin{subfigure}[b]{0.49\textwidth}
    \begin{tikzpicture}
      \begin{semilogxaxis}[
            xlabel=Gibbs update,
            xmax=1000,
            xmin=1,
        ]
        \confplot{Naive}{classic}{recall}{classical}{style0}
        \confplot{Naive}{dwave_8}{recall}{D-Wave, $T=8$}{style1}
        \confplot{Naive}{dwave_16}{recall}{D-Wave, $T=16$}{style2}
        \confplot{Naive}{dwave_32}{recall}{D-Wave, $T=32$}{style3}
        \confplot{Naive}{dwave_64}{recall}{D-Wave, $T=64$}{style4}
        \legend{}
      \end{semilogxaxis}
    \end{tikzpicture}
    \caption{Recall}\label{fig:classic_v_dw_recall}
  \end{subfigure}
  \begin{subfigure}[b]{0.49\textwidth}
    \begin{tikzpicture}
      \begin{semilogxaxis}[
            xlabel=Gibbs update,
            xmax=1000,
            xmin=1,
            ymax=0.3,
        ]
        \confplot{Naive}{classic}{l2}{classical}{style0}
        \confplot{Naive}{dwave_8}{l2}{D-Wave, $T=8$}{style1}
        \confplot{Naive}{dwave_16}{l2}{D-Wave, $T=16$}{style2}
        \confplot{Naive}{dwave_32}{l2}{D-Wave, $T=32$}{style3}
        \confplot{Naive}{dwave_64}{l2}{D-Wave, $T=64$}{style4}
        \legend{}
      \end{semilogxaxis}
    \end{tikzpicture}
    \caption{Positive-case distribution distance}\label{fig:classic_v_dw_l2}
  \end{subfigure}
  \begin{subfigure}[b]{0.49\textwidth}
    \begin{tikzpicture}
      \begin{semilogxaxis}[
            xlabel=Gibbs update,
            xmax=1000,
            xmin=1,
        ]
        \confplot{Naive}{classic}{MADneg}{classical}{style0}
        \confplot{Naive}{dwave_8}{MADneg}{D-Wave, $T=8$}{style1}
        \confplot{Naive}{dwave_16}{MADneg}{D-Wave, $T=16$}{style2}
        \confplot{Naive}{dwave_32}{MADneg}{D-Wave, $T=32$}{style3}
        \confplot{Naive}{dwave_64}{MADneg}{D-Wave, $T=64$}{style4}
        \legend{}
      \end{semilogxaxis}
    \end{tikzpicture}
    \caption{Negative-case mean edit distance}\label{fig:classic_v_dw_mad}
  \end{subfigure}
  \caption{Characteristics of the samples generated by Naively trained RBMs after a number of Gibbs updates with different initializations. For most characteristics, the differences due to the different initializations remain significant even after \num{1000} Gibbs updates.}
  \label{fig:classic_v_dw}
  \end{center}
\end{figure}

%% file: figures/longexperiments.tex
\begin{figure}
  \begin{center}
  \ref*{long_legend}
  \begin{subfigure}[b]{0.49\textwidth}
    \begin{tikzpicture}
      \begin{semilogxaxis}[
            xlabel=Gibbs update,
            xmax=100000,
            xmin=1,
            legend pos=south east,
            legend to name=long_legend
        ]
        \confplot[inference_long]{Naive}{classic}{precision}{classical}{style0}
        \confplot[inference_long]{Naive}{dwave_8}{precision}{D-Wave, $T=8$}{style1}
        \confplot[inference_long]{Naive}{dwave_32}{precision}{D-Wave, $T=32$}{style3}
        \confplot[inference_long]{Naive}{hybrid_8}{precision}{Hybrid, $T=8$}{style2}
        \confplot[inference_long]{Naive}{hybrid_32}{precision}{Hybrid, $T=32$}{style4}
      \end{semilogxaxis}
    \end{tikzpicture}
    \caption{Precision}\label{fig:long_precision}
  \end{subfigure}
  \begin{subfigure}[b]{0.49\textwidth}
    \begin{tikzpicture}
      \begin{semilogxaxis}[
            xlabel=Gibbs update,
            xmax=100000,
            xmin=1,
        ]
        \confplot[inference_long]{Naive}{classic}{recall}{classical}{style0}
        \confplot[inference_long]{Naive}{dwave_8}{recall}{D-Wave, $T=8$}{style1}
        \confplot[inference_long]{Naive}{dwave_32}{recall}{D-Wave, $T=32$}{style3}
        \confplot[inference_long]{Naive}{hybrid_8}{recall}{Hybrid, $T=8$}{style2}
        \confplot[inference_long]{Naive}{hybrid_32}{recall}{Hybrid, $T=32$}{style4}
        \legend{}
      \end{semilogxaxis}
    \end{tikzpicture}
    \caption{Recall}\label{fig:long_recall}
  \end{subfigure}
  \begin{subfigure}[b]{0.49\textwidth}
    \begin{tikzpicture}
      \begin{semilogxaxis}[
            xlabel=Gibbs update,
            xmax=100000,
            xmin=1,
            ymax=0.3,
        ]
        \confplot[inference_long]{Naive}{classic}{l2}{classical}{style0}
        \confplot[inference_long]{Naive}{dwave_8}{l2}{D-Wave, $T=8$}{style1}
        \confplot[inference_long]{Naive}{dwave_32}{l2}{D-Wave, $T=32$}{style3}
        \confplot[inference_long]{Naive}{hybrid_8}{l2}{Hybrid, $T=8$}{style2}
        \confplot[inference_long]{Naive}{hybrid_32}{l2}{Hybrid, $T=32$}{style4}
        \legend{}
      \end{semilogxaxis}
    \end{tikzpicture}
    \caption{Positive-case distribution distance}\label{fig:long_l2}
  \end{subfigure}
  \begin{subfigure}[b]{0.49\textwidth}
    \begin{tikzpicture}
      \begin{semilogxaxis}[
            xlabel=Gibbs update,
            xmax=100000,
            xmin=1,
        ]
        \confplot[inference_long]{Naive}{classic}{MADneg}{classical}{style0}
        \confplot[inference_long]{Naive}{dwave_8}{MADneg}{D-Wave, $T=8$}{style1}
        \confplot[inference_long]{Naive}{dwave_32}{MADneg}{D-Wave, $T=32$}{style3}
        \confplot[inference_long]{Naive}{hybrid_8}{MADneg}{Hybrid, $T=8$}{style2}
        \confplot[inference_long]{Naive}{hybrid_32}{MADneg}{Hybrid, $T=32$}{style4}
        \legend{}
      \end{semilogxaxis}
    \end{tikzpicture}
    \caption{Negative-case mean edit distance}\label{fig:long_mad}
  \end{subfigure}
  \caption{Sample statistic development for a larger number of Gibbs updates for naively-trained RBMs. Note that initialization affects the results even after \num{100000} Gibbs updates. For example, the mean absolute distance of the negative samples to the closest positive sample (\ref{fig:long_mad}) has reached a plateau at substantially higher figure than any of the other initializations, indicating that some of the Markov chains have been stuck in states with negative images.}
  \label{fig:long}
  \end{center}
\end{figure}

%% file: figures/top10.tex
\begin{figure}
  \begin{center}
    \begin{tikzpicture}
      \begin{semilogyaxis}[
        ytick={1000,2000,3000,4000,5000,6000,7000,8000,9000},
        yticklabels={1000,2000,3000,4000,5000,6000,7000,8000,9000},
        xtick={0,1,2,3,4,5,6,7,8},
        xticklabels={classical,\text{D-Wave, $T=64$},\text{D-Wave, $T=32$},\text{D-Wave, $T=16$},\text{D-Wave, $T=8$},\text{Hybrid, $T=64$},\text{Hybrid, $T=32$},\text{Hybrid, $T=16$},\text{Hybrid, $T=8$}},
        xticklabel style={rotate=90},
        xmin=-1,
        ymin=900,
        ymax=9000,
        height=7cm,
      ]
      \addplot[only marks,mark=*,opacity=0.4] table[x=position, y=top10, col sep=comma] {data/top.csv};
      \end{semilogyaxis}
    \end{tikzpicture}
  \caption{The figure shows the number of times each model generated one of its top-10 most frequently generated positive images. It demonstrates whether the model tends to favor some positive images over others, deviating from the uniform distribution. Only naively-trained RBMs are shown. Each point corresponds to one RBM instance depending on the initialization. If the \num{40000} samples came from a uniform distribution of all \num{8190} positive images, the expected sum of the top-10 most frequent occurrences would be below \num{150}. In the same scenario with \num{819} positive images (modeling recall of about \num{0.1}), the expected sum would be below \num{750}. The graph shows that with lower temperatures $T$, D-Wave samples concentrate around some high-probability states. The hybrid strategy alleviates some of this concentration.}
  \label{fig:top10}
  \end{center}
\end{figure}

%% file: figures/classic_dw_hybrid.tex
\begin{figure}
  \begin{center}
  \ref*{classic_dw_hybrid_legend}
  \begin{subfigure}[b]{0.49\textwidth}
    \begin{tikzpicture}
      \begin{semilogxaxis}[
            xlabel=Gibbs update,
            xmax=1000,
            xmin=1,
            legend to name=classic_dw_hybrid_legend
        ]
        \confplot{Naive}{classic}{precision}{classical}{style0}
        \confplot{Naive}{dwave_8}{precision}{D-Wave, $T=8$}{style1}
        \confplot{Naive}{hybrid_8}{precision}{Hybrid, $T=8$}{style2}
        \confplot{Naive}{dwave_32}{precision}{D-Wave, $T=32$}{style3}
        \confplot{Naive}{hybrid_32}{precision}{Hybrid, $T=32$}{style4}
      \end{semilogxaxis}
    \end{tikzpicture}
    \caption{Precision}\label{fig:classic_dw_hybrid_precision}
  \end{subfigure}
  \begin{subfigure}[b]{0.49\textwidth}
    \begin{tikzpicture}
      \begin{semilogxaxis}[
            xlabel=Gibbs update,
            xmax=1000,
            xmin=1,
        ]
        \confplot{Naive}{classic}{recall}{classical}{style0}
        \confplot{Naive}{dwave_8}{recall}{D-Wave, $T=8$}{style1}
        \confplot{Naive}{hybrid_8}{recall}{Hybrid, $T=8$}{style2}
        \confplot{Naive}{dwave_32}{recall}{D-Wave, $T=32$}{style3}
        \confplot{Naive}{hybrid_32}{recall}{Hybrid, $T=32$}{style4}
        \legend{}
      \end{semilogxaxis}
    \end{tikzpicture}
    \caption{Recall}\label{fig:classic_dw_hybrid_recall}
  \end{subfigure}
  \caption{The effect of hybrid initialization at $T=8$ and $T=32$. Mixing the classical and D-Wave initialization strategies alleviates some of the weaknesses of both. Hybrid initialization at $T=32$ achieves the highest peak recall, and preserves this advantage over classical initialization at least up to \num{1000} Gibbs updates.}
  \label{fig:classic_dw_hybrid}
  \end{center}
\end{figure}

%% file: figures/shifter_naive.tex
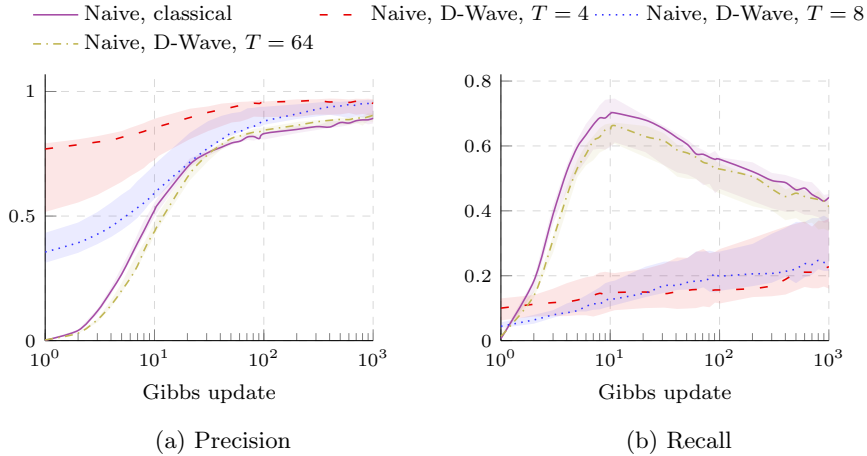
\begin{figure}
  \begin{center}
  \ref*{shifter_naive_legend}
  \begin{subfigure}[b]{0.49\textwidth}
    \begin{tikzpicture}
      \begin{semilogxaxis}[
            xlabel=Gibbs update,
            xmax=1000,
            xmin=1,
            legend pos=south east,
            legend to name=shifter_naive_legend
        ]
        \confplot[shifter_inference]{Naive}{classic}{precision}{Naive, classical}{style0}
%        \confplot[shifter_inference]{Naive}{dwave_2}{precision}{Naive, D-Wave, $T=2$}{style1}
        \confplot[shifter_inference]{Naive}{dwave_4}{precision}{Naive, D-Wave, $T=4$}{style2}
        \confplot[shifter_inference]{Naive}{dwave_8}{precision}{Naive, D-Wave, $T=8$}{style3}
        \confplot[shifter_inference]{Naive}{dwave_64}{precision}{Naive, D-Wave, $T=64$}{style4}
        
      \end{semilogxaxis}
    \end{tikzpicture}
    \caption{Precision}\label{fig:shifter_naive_precision}
  \end{subfigure}
  \begin{subfigure}[b]{0.49\textwidth}
    \begin{tikzpicture}
      \begin{semilogxaxis}[
            xlabel=Gibbs update,
            %ylabel=Recall,
            xmax=1000,
            xmin=1,
        ]
        \confplot[shifter_inference]{Naive}{classic}{recall}{Naive, classical}{style0}
%        \confplot[shifter_inference]{Naive}{dwave_2}{recall}{Naive, D-Wave, $T=2$}{style1}
        \confplot[shifter_inference]{Naive}{dwave_4}{recall}{Naive, D-Wave, $T=4$}{style2}
        \confplot[shifter_inference]{Naive}{dwave_8}{recall}{Naive, D-Wave, $T=8$}{style3}
        \confplot[shifter_inference]{Naive}{dwave_64}{recall}{Naive, D-Wave, $T=64$}{style4}
        \legend{}
      \end{semilogxaxis}
    \end{tikzpicture}
    \caption{Recall}\label{fig:shifter_naive_recall}
  \end{subfigure}

  \caption{D-Wave initialization at $T=4$ for the naively trained RBM reaches high precision instantly, i.e. \dwave is able to generate mostly positive examples even without additional Gibbs updates. However, concentrating the initial values around the high-probability states has a negative effect on recall, which remains noticeable even after \num{1000} Gibbs updates.}
  \label{fig:shifter_naive}
  \end{center}
\end{figure}

%% file: figures/shifter_naive_small_T.tex
\begin{figure}
  \begin{center}
  \ref*{shifter_naive_small_t_legend}
  \begin{subfigure}[b]{0.49\textwidth}
    \begin{tikzpicture}
      \begin{semilogxaxis}[
            xlabel=Gibbs update,
            xmax=1000,
            xmin=1,
            legend pos=south east,
            legend to name=shifter_naive_small_t_legend
        ]
        \confplot[shifter_inference]{Naive}{dwave_2}{precision}{Naive, D-Wave, $T=2$}{style0}
        \confplot[shifter_inference]{Naive}{dwave_3}{precision}{Naive, D-Wave, $T=3$}{style1}
        \confplot[shifter_inference]{Naive}{dwave_4}{precision}{Naive, D-Wave, $T=4$}{style2}
        \confplot[shifter_inference]{Naive}{dwave_5}{precision}{Naive, D-Wave, $T=5$}{style3}
        \confplot[shifter_inference]{Naive}{dwave_6}{precision}{Naive, D-Wave, $T=6$}{style4}
        
      \end{semilogxaxis}
    \end{tikzpicture}
    \caption{Precision}\label{fig:shifter_naive_small_t_precision}
  \end{subfigure}
  \begin{subfigure}[b]{0.49\textwidth}
    \begin{tikzpicture}
      \begin{semilogxaxis}[
            xlabel=Gibbs update,
            xmax=1000,
            xmin=1,
        ]
        \confplot[shifter_inference]{Naive}{dwave_2}{recall}{Naive, D-Wave, $T=2$}{style0}
        \confplot[shifter_inference]{Naive}{dwave_3}{recall}{Naive, D-Wave, $T=3$}{style1}
        \confplot[shifter_inference]{Naive}{dwave_4}{recall}{Naive, D-Wave, $T=4$}{style2}
        \confplot[shifter_inference]{Naive}{dwave_5}{recall}{Naive, D-Wave, $T=5$}{style3}
        \confplot[shifter_inference]{Naive}{dwave_6}{recall}{Naive, D-Wave, $T=6$}{style4}
        \legend{}
      \end{semilogxaxis}
    \end{tikzpicture}
    \caption{Recall}\label{fig:shifter_naive_small_t_recall}
  \end{subfigure}

  \caption{The properties of the samples generated with D-Wave depend substantially on the temperature used. However, this hyperparameter cannot always be viewed in isolation from the physical device's idiosyncrasies. For example, it could be expected that precision increases as the temperature $T$ decreases, since more of the samples should come from the low energy states. However, at $T=2$ the coupler strengths used to ensure the consistency of physical qubit chains forming the logical qubits start to be similar to coupler strengths required to represent the RBM weights. This introduces inconsistencies that lower the quality of the samples.}
  \label{fig:shifter_naive_small_t}
  \end{center}
\end{figure}
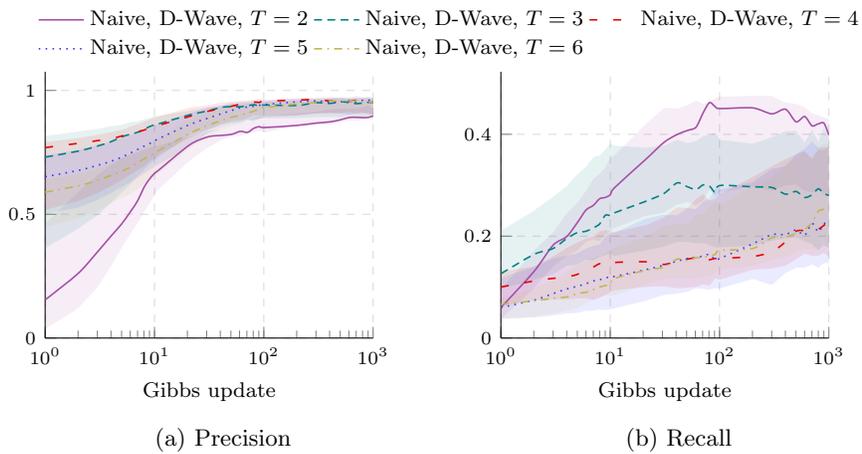

%% file: figures/shifter_cd1.tex
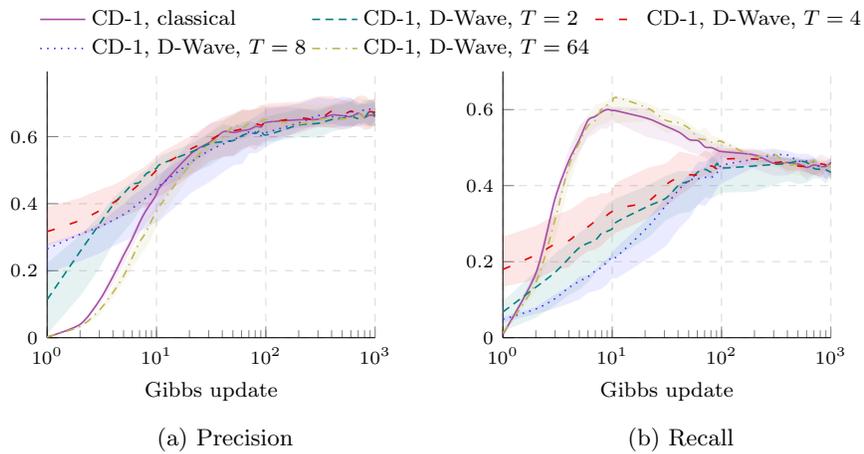
\begin{figure}
  \begin{center}
  \ref*{shifter_cd1_legend}
  \begin{subfigure}[b]{0.49\textwidth}
    \begin{tikzpicture}
      \begin{semilogxaxis}[
            xlabel=Gibbs update,
            xmax=1000,
            xmin=1,
            legend pos=south east,
            legend to name=shifter_cd1_legend
        ]
        \confplot[shifter_inference]{CD1}{classic}{precision}{CD-1, classical}{style0}
        \confplot[shifter_inference]{CD1}{dwave_2}{precision}{CD-1, D-Wave, $T=2$}{style1}
        \confplot[shifter_inference]{CD1}{dwave_4}{precision}{CD-1, D-Wave, $T=4$}{style2}
        \confplot[shifter_inference]{CD1}{dwave_8}{precision}{CD-1, D-Wave, $T=8$}{style3}
        \confplot[shifter_inference]{CD1}{dwave_64}{precision}{CD-1, D-Wave, $T=64$}{style4}
        
      \end{semilogxaxis}
    \end{tikzpicture}
    \caption{Precision}\label{fig:shifter_cd1_precision}
  \end{subfigure}
  \begin{subfigure}[b]{0.49\textwidth}
    \begin{tikzpicture}
      \begin{semilogxaxis}[
            xlabel=Gibbs update,
            xmax=1000,
            xmin=1,
        ]
        \confplot[shifter_inference]{CD1}{classic}{recall}{CD-1, classical}{style0}
        \confplot[shifter_inference]{CD1}{dwave_2}{recall}{CD-1, D-Wave, $T=2$}{style1}
        \confplot[shifter_inference]{CD1}{dwave_4}{recall}{CD-1, D-Wave, $T=4$}{style2}
        \confplot[shifter_inference]{CD1}{dwave_8}{recall}{CD-1, D-Wave, $T=8$}{style3}
        \confplot[shifter_inference]{CD1}{dwave_64}{recall}{CD-1, D-Wave, $T=64$}{style4}
        \legend{}
      \end{semilogxaxis}
    \end{tikzpicture}
    \caption{Recall}\label{fig:shifter_cd1_recall}
  \end{subfigure}

  \caption{CD-1 inference with various initializations for the Shifter problem. For a very small number of Gibbs updates, CD-1 trained models benefit from initialization closer to positive examples, i.e., D-Wave initialization with low temperature $T$. More random initialization (classical and high-temperature D-Wave) leads to higher maximum recall. The differences are not permanent and all initialization methods perform similarly after about \num{100} Gibbs update steps.}
  \label{fig:shifter_cd1}
  \end{center}
\end{figure}